\documentclass[preprint,onecolumn,nofootinbib]{revtex4}
\pdfoutput=1
\usepackage[colorlinks=true,linkcolor=blue,urlcolor=blue,filecolor=black,citecolor=red,pdfstartview=FitV,pdftitle={},pdfsubject={},pdfkeywords={},pdfpagemode=None,bookmarksopen=true]{hyperref}
\usepackage{graphicx}%Include figure files
\usepackage{amsmath}
\usepackage{amsfonts}
\usepackage{amssymb,ulem}
\usepackage{dcolumn,booktabs}%
\usepackage{color,xcolor}%
\usepackage{CJK}
\usepackage{subfigure}

\usepackage{dcolumn}% Align table columns on decimal pointl
\usepackage{float}
\usepackage{multirow}% apply to table
\usepackage{graphicx}
\newcommand{\f}{\begin{equation}}
\newcommand{\ff}{\end{equation}}
\newcommand{\fa}{\begin{eqnarray}}
\newcommand{\ffa}{\end{eqnarray}}
\begin{document}
\title{Probing loop quantum effects through solar system experiments: observational signatures and parameter constraints}

\author{Wen-Juan Ai$^{1}$}
\thanks{wenjuanai2025@163.com}
\author{Ruo-Ting Chen$^{1}$}
\thanks{ruotingchen@163.com}
\author{Li-Gang Zhu$^{1}$}
\thanks{zlgoupao@163.com}
\author{Jian-Pin Wu$^{1}$}
\thanks{jianpinwu@yzu.edu.cn, corresponding author} 
\affiliation{
  $^1$ Center for Gravitation and Cosmology, College of Physical
  Science and Technology, Yangzhou University, Yangzhou 225009,
  China}

\begin{abstract}

This study investigates quantum gravity effects within the framework of an effective loop quantum gravity black hole (LQG-BH) model parameterized by $\zeta$, utilizing precision measurements from solar system experiments and astrophysical observations. We analyze three classical tests of general relativity (GR): (1) Light deflection constrained by very long baseline interferometry (VLBI) observations of quasar radio signals, (2) Shapiro time delay measurements from the Cassini mission, and (3) Mercury's perihelion precession determined by MESSENGER mission data. Additionally, we extend our analysis to Earth-orbiting LAGEOS satellites and the relativistic trajectory of the S2 star orbiting the Galactic Center supermassive BH Sagittarius $\rm{A}^*$ (Sgr $\rm{A}^{*}$). Our multi-probe approach reveals that the tightest constraint on the LQG parameter comes from Mercury's perihelion precession, yielding an upper bound $\zeta \lesssim 10^{-2}$. These results establish new observational benchmarks for probing quantum gravity effects.

\end{abstract}

\maketitle
\tableofcontents

\section{Introduction}

Over the past century, Einstein’s general relativity (GR) has not only revolutionized our understanding of spacetime and gravity but has also triumphantly survived the most rigorous and precise observational tests across both weak-field and strong-field regimes. In weak-field gravity, GR’s predictions have been validated through precise measurements of astrophysical phenomena such as the perihelion advance of planetary orbits \cite{Will:2014kxa,Park:2017zgd}, the deflection of light \cite{Fomalont:2009zg}, and the Shapiro time delay \cite{Bertotti:2003rm}. In the strong-field regime, GR has been tested against extreme astrophysical systems, including binary pulsar dynamics \cite{Stairs:2003eg,EventHorizonTelescope:2019dse}, black hole (BH) shadow imaging \cite{EventHorizonTelescope:2020qrl}, and gravitational wave (GW) detections from merging compact objects \cite{LIGOScientific:2016aoc}. Remarkably, GR’s predictions remain consistent with observations at the current sensitivity levels — a testament to the theory’s enduring robustness and its foundational role in modern physics.

Despite its remarkable theoretical and empirical robustness, GR faces unresolved issues that demand beyond-standard frameworks. These challenges include the theoretical limitations and observational anomalies. Theoretically, GR predicts spacetime singularities at cosmological origins \cite{Borde:1993xh} and BH centers \cite{Hawking:1973uf}, where curvature divergences terminate predictability. In addition, no known formalism consistently unifies GR with quantum mechanics \cite{Adler:2010wf,Ng:2003jk}, leaving quantum gravity as an open frontier. Observationally, dark matter halos and dark energy, empirically required by $\Lambda$CDM cosmology, lack fundamental justification within GR. Potential tensions in extreme environments, e.g., BH mergers, early-universe physics, may hint at beyond-GR effects.

One of the most effective ways to address these anomalies is to develop a consistent quantum theory of gravity. Among quantum gravity candidates, loop quantum gravity (LQG) provides a non-perturbative, background-independent framework \cite{Thiemann:2007pyv,Smolin:2006pa,Han:2005km}. The cosmological implementation of LQG, known as loop quantum cosmology (LQC), demonstrates singularity resolution by incorporating two key quantum corrections: the inverse volume correction and the holonomy correction \cite{Bojowald:2001xe,Ashtekar:2006rx,Ashtekar:2006uz,Ashtekar:2006wn,Ashtekar:2003hd,Bojowald:2005epg,Ashtekar:2011ni}. This framework replaces the Big Bang singularity with a nonsingular quantum bounce \cite{Ashtekar:2011ni}, which then evolves into the current state of the universe \cite{Ashtekar:2011ni,Ashtekar:2015dja}. The LQC paradigm naturally extends to spherically symmetric BHs, yielding LQG-BHs. For technical details on LQG-BH construction, see \cite{Chiou:2008nm,Chiou:2008eg,Boehmer:2007ket}; comprehensive reviews in \cite{Perez:2017cmj,Modesto:2005zm,Modesto:2008im}. In LQG-BHs, the singularity is resolved, and a quantum transition surface typically bridges the trapped and anti-trapped regions \cite{Ashtekar:2005qt,Zhang:2023yps,Zhang:2023okw, Zhang:2024ney}.

In recent decades, cosmology has achieved remarkable maturity, driven partly by increasingly precise Cosmic Microwave Background (CMB) measurements. Studies indicate that the pre-inflationary dynamics of LQC imprint deviations from near-scale invariance in primordial power spectra \cite{Ashtekar:2020gec,Ashtekar:2021izi,Agullo:2020fbw,Agullo:2021oqk,Zhang:2007bi,Li:2011ac,deBlas:2016puz,Martin-Benito:2021szh}. Moreover, GW detections from binary mergers \cite{LIGOScientific:2016aoc,LIGOScientific:2016lio,LIGOScientific:2016sjg} and Event Horizon Telescope (EHT) imaging of supermassive black holes (SMBH) $\mathrm{M87^*}$ and $\mathrm{Sgr\ A^{*}}$ \cite{EventHorizonTelescope:2019dse,EventHorizonTelescope:2019ths,EventHorizonTelescope:2022xnr,EventHorizonTelescope:2022xqj} have enabled probes of quantum gravity in strong-field regimes. Consequently, most phenomenological studies of LQG have focused on high-energy and strong-curvature scenarios, exploring imprints through quasi-normal mode (QNM) spectrum \cite{Konoplya:2024lch,Skvortsova:2024msa,Zhang:2024svj,Fu:2023drp,Gong:2023ghh,Zhu:2024wic,Shao:2023qlt}, photon rings, shadow morphology \cite{Liu:2024soc, Liu:2024wal, Xamidov:2025oqx}, spinning particle dynamics \cite{Du:2024ujg}, accretion disk structures \cite{Shu:2024tut, Chen:2025ifv,Zhang:2023okw,Yang:2022btw}, and GW radiations from periodic orbits \cite{Tu:2023xab} or extreme mass-ratio inspirals \cite{Fu:2024cfk,Zi:2024jla}, etc. In contrast, this work shifts focus to precision weak field tests--a domain where solar system experiments have proven exceptionally effective for constraining gravitational theories \cite{Zhu:2020tcf,Liu:2022qiz,Chen:2023bao,Zhu:2024oxz}. Although strong-curvature regimes remain important, the unparalleled precision of local gravitational measurements offers complementary constraints on LQG effects.

In this paper, we will probe quantum gravity effects through solar system experiments in a LQG-inspired BH spacetime featuring double horizons, parameterized by dimensionless deformation $\zeta$ \cite{Zhang:2024khj}. The paper is organized as follows. Section \ref{bhsolution} provides a detailed geometric characterization of the effective LQG-BH spacetime, with particular emphasis on the geodesic motion of particles in its exterior spacetime. In Section \ref{constrain}, the classical GR experiments are employed to probe the effects of the quantum gravity effect. Section \ref{conclusion} summarizes the key findings and describes prospective directions for future investigations. Throughout this work, we adopt geometrized Planck units unless otherwise specified. When using experimental data, the International System of Units (SI) is restored for calculations.

%%%%%%%%%%%%%%%%%%%%%%%%%%%%%%%%%%%%%%%%%%

\section{Test Particle Dynamics in an Effective LQG-BH Spacetime}\label{bhsolution}

In this section, we first outline the effective LQG-BH spacetime \cite{Zhang:2024khj}, then derive the equations of motion (EOMs) for test particles orbiting this quantum-corrected geometry.

\subsection{An effective LQG-BH spacetime}\label{sec-BH}

The LQG-BH geometry, originally proposed in \cite{Zhang:2024khj}, is given by:
\begin{eqnarray}
\label{metric}
\mathrm{d}s^{2}=-{f\left ( r \right )}\mathrm{d}t^{2}+\frac{1}{f\left ( r \right ) }\mathrm{d}r^{2}+r^{2}\left (\mathrm{d}\theta ^{2}+\sin^{2}\theta\mathrm{d}\phi^{2}\right ) \,,
\end{eqnarray}
where the metric function $f(r)$ takes the explicit form
\begin{eqnarray}
f\left ( r \right )=1-\frac{2M}{r}+\frac{M^{2}\zeta ^{2}}{r^{2}}\left ( 1-\frac{2M}{r}  \right ) ^{2} \,.
\label{equ:II.A-(1)}
\end{eqnarray}
The dimensionless parameter $\zeta$ encodes quantum gravitational corrections and is fundamentally defined as:
\begin{equation}
\zeta = \frac{\gamma \sqrt{A}}{M}\,
\label{zeta_M}
\end{equation}
where $M$ is the BH mass, $\gamma$ denotes the Barbero-Immirzi (BI) parameter of LQG, and $A$ represents the minimal area gap from LQG's holonomy quantization. Specifically, $A = 4\sqrt{3} \pi \gamma \ell_{\mathrm{P}}^{2}$ corresponds to the smallest non-zero eigenvalue of LQG's area operator \cite{Ashtekar:2006wn}, with $\ell_{\mathrm{P}}$ being the Planck length. This parameter $\zeta$ determines the onset of quantum gravity effects, where the classical singularity ($A \to 0$) is recovered as $\zeta \to 0$. Since the BI parameter $\gamma$ currently lacks first-principles determination in current LQG frameworks \cite{Meissner:2004ju,Domagala:2004jt}, we treat $\zeta$ as an effective dimensionless free parameter in our phenomenological approach. This approach allows systematic investigation of LQG-induced modifications to BH.

\begin {figure}[h]
    \centering
    \begin {minipage} {0.48\textwidth}
        \centering
        \includegraphics[
   width = \linewidth] {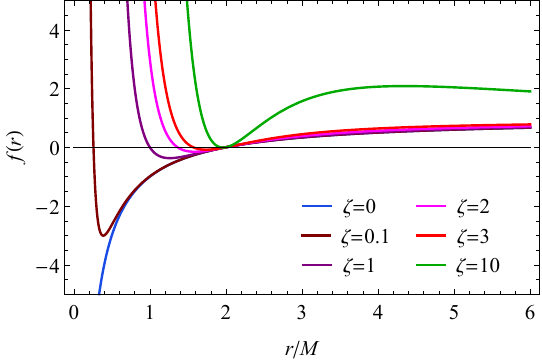}
    \end {minipage}
    \hfill
    \begin {minipage} {0.48\textwidth}
        \centering
        \includegraphics[
   width = \linewidth] {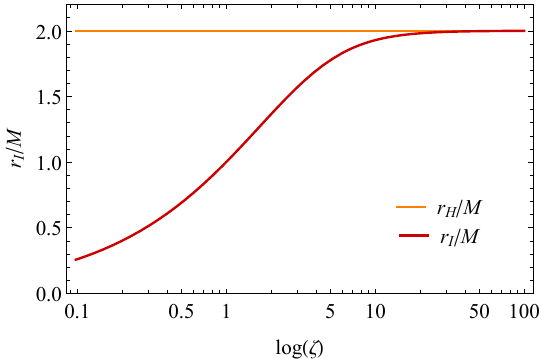}
    \end {minipage}
    \caption {Left: The metric function $f(r)$ for varying values of the LQG-corrected parameter $\zeta$. The blue curve corresponds to the Schwarzschild case ($\zeta=0$). 
    Right: The inner horizon radius $r_{\mathrm{I}}$ as a function of $\zeta$. The orange line indicates the event horizon $r_{\mathrm{H}}$ for reference.}
    \label {fig1}
\end {figure}
To analyze the LQG-corrected BH properties, we first examine the metric function $f(r)$ under varying quantum parameter $\zeta$, as illustrated in the left panel of Fig. \ref{fig1}. The equation $g^{rr}=0$, which is equivalent to $f(r)=0$, admits two real roots: one corresponding to the event horizon at $r_{\mathrm{H}}=2 M$, and the other representing the inner horizon, given by  
\begin{eqnarray}
r_{\mathrm{I}} = \frac{M \zeta^{4/3}}{3^{1/3}\left(-9 + \sqrt{81 + 3 \thinspace \zeta^{2}}\right)^{1/3}}-\frac{M \zeta^{2/3} \left(-9 + \sqrt{81 + 3 \thinspace \zeta^{2}}\right)^{1/3}}{3^{2/3}} \,.
\label{ri}
\end{eqnarray} 
The right panel of Fig. \ref{fig1} illustrates the functional dependence of $r_{\mathrm{I}}$ on $\zeta$. It is evident that in the classical limit of $\zeta$, the spacetime reduces to the Schwarzschild scenario, with the inner horizon $r_{\mathrm{I}}$ collapsing to the singularity at $r=0$. In the finite regime of $\zeta$, i.e., $0<\zeta<\infty$, the inner horizon radius $r_{\mathrm{I}}$ monotonically increases with $\zeta$, maintaining a hierarchy $r_{\mathrm{I}}<r_{\mathrm{H}}$ throughout (see the right panel of Fig. \ref{fig1}). Expanding Eq. \eqref{ri} in the limit of $\zeta \to \infty $ yields
\begin{eqnarray}
r_{\mathrm{I}} = 2M - \frac{8M}{\zeta^2}+\mathcal{O}\left( \zeta^{-3} \right)\,,
\end{eqnarray} 
demonstrating that $r_{\mathrm{I}}$ asymptotically approaches $r_{\rm{H}}$ while preserving for any finite $\zeta$. This behavior is further corroborated by the left panel of Fig. \ref{fig1}. The introduction of quantum gravity effects resolves the classical Schwarzschild singularity in this spacetime, replacing it with a transition region that connects a BH to a white hole \cite{Zhang:2024khj}. This region features a bounce surface located within the range $0<r_{\mathrm{B}}<r_{\mathrm{I}}$, where $r_{\mathrm{B}}$ denotes the bounce radius. Such a causal structure aligns closely with those in other LQG-BH models \cite{Munch:2021oqn, Lewandowski:2022zce}.

%%%%%%%%%%%%%%%%%%%%%%%%%%%%%%
\subsection{Test particle dynamics}

The Lagrangian governing test particle motion can be expressed as:
\begin{eqnarray}
\mathcal{L}\left ( x^{\mu},\dot{x} ^{\mu} \right ) =\frac{1}{2}  g_{\mu \nu} \dot{x}^{\mu} \dot{x}^{\nu}\,,
\label{eq:Lagrangian}
\end{eqnarray}
where the overdot denotes differentiation with respect to the affine parameter $\lambda$ along geodesics.
Substituting the spherically symmetric metric (Eq.~\eqref{metric}) into Eq.~\eqref{eq:Lagrangian}, we obtain the explicit form:
%%%%%
\begin{eqnarray}
\mathcal{L}\left ( x^{\mu},\dot{x} ^{\mu} \right ) =\frac{1}{2} \left ( -f(r)\dot{t}^{2}+ f(r)^{-1}\dot{r}^{2}+r^{2}\dot{\theta }^{2}+r^{2}\sin^{2}\theta  \dot{\phi }^{2}\right )\,, 
\label{Lagrangian equation}
\end{eqnarray}
The geodesic equations are derived via the Euler-Lagrange formalism:
\begin{eqnarray}
\frac{\mathrm{d} }{\mathrm{d} \lambda} \frac{\partial \mathcal{L}}{\partial \dot{x^{\mu }} } -\frac{\partial \mathcal{L}}{\partial x^{\mu}}=0\,. 
\label{Euler-Lagrange equation}
\end{eqnarray}

The Euler-Lagrange equation explicitly reveals two conserved quantities associated with spacetime symmetries, namely the energy $E$ and the angular momentum $J$:
\begin{eqnarray}
\frac{\partial\mathcal{L} }{\partial\dot{t}} &=& -f(r)\dot{t}\equiv-E\,,
\label{energy}  \\
\frac{\partial\mathcal{L}}{\partial\dot{\phi}} &=& r^{2}\sin^{2}\theta\dot{\phi}\equiv J \,.
\label{Amomentum}
\end{eqnarray}
These conservation laws originate from the spacetime's stationarity (time-translation invariance) and axisymmetry (rotational invariance about the polar axis), respectively.

For timelike ($\eta=1$) or null ($\eta=0$) geodesics, the four-velocity satisfies:
\begin{eqnarray}
g_{\mu \nu} \frac{d x^{\mu}}{d \lambda} \frac{d x^{\nu}}  {d \lambda}=-\eta \,,
\label{eq:Normalization} 
\end{eqnarray}
where $\eta=1$ corresponds to massive particles and $\eta=0$ to massless particles.
Constraining the motion to the equatorial plane ($\theta=\frac{\pi}{2}$, $\dot{\theta}=0$) and combining Eq.~\eqref{eq:Normalization} with the conserved quantities from Eqs. \eqref{energy} and \eqref{Amomentum}, we derive the radial equation of motion:
%%%%%
\begin{eqnarray}
\left(\frac{d r}{d \lambda}\right)^{2}=E^{2}-f\left ( r \right ) \left ( \eta +\frac{J^{2}}{r^{2}}  \right ) \,,
\label{Eq:rE}
\end{eqnarray}
accompanied by the temporal and azimuthal evolution equations:
%%%%%
\begin{eqnarray}
\frac{\mathrm{d} t}{\mathrm{d} \lambda } &=& \frac{{E}}{f\left(r\right)}\,,
\label{Eq:tE}  \\
\frac{\mathrm{d} \phi}{\mathrm{d} \lambda} &=& \frac{{J}}{r^{2}}  \,.
\label{Eq:phiJ}
 \end{eqnarray}
These equations provides the foundation for calculating key observational effects in this LQG-corrected geometry including the light deflection angle, Shapiro time delay, and periastron precession.

%%%%%%%%%%%%%%%%%%%%%%%%%%%%%%%%%%%%%%%%%%
\section{Constraints on quantum parameter}\label{constrain}

This section derives quantitative constraints on LQG-corrected parameter $\zeta$ through solar system tests. Using the relativistic framework developed in previous sections, we calculate three classical gravitational effects, including the light deflection angle, Shapiro time delay, and periastron precession within the effective LQG-corrected BH spacetime.

\subsection{Deflection of light}\label{deflection}

Consider a light ray propagating along a null geodesic ($\eta=0$) in the solar gravitational field, originating from spatial infinity, reaching a closest approach at radial coordinate $r_0$, and escaping back to infinity. The angular deflection per unit radial displacement, derived from the geodesic Eqs. \eqref{Eq:rE} and \eqref{Eq:phiJ}, is governed by
%%%%%
\begin{eqnarray}
\frac{\mathrm{d}\phi }{\mathrm{d}r}= \pm \left ( \frac{r^{4}}{b^{2}}-f\left ( r \right )r^{2}\right )^{-\frac{1}{2} }\,, 
\label{drdphi}
\end{eqnarray} 
where the impact parameter $b\equiv J/E$ characterizes the trajectory's initial conditions. Geometrically, $b$ represents the perpendicular distance between the undeflected light path in the absence of gravity and the Sun’s centerline. The $±$ sign corresponds to the outgoing ($+$) and ingoing ($-$) trajectory segments during its gravitational encounter.

At the closest approach $r_0$, the radial turning point condition
\begin{eqnarray}
    \frac{\mathrm{d}r }{\mathrm{d}\phi }\mid _{r=r_{0}}=0\,,
    \label{TPcondition}
\end{eqnarray}
directly follows from \eqref{drdphi}. This imposes a geometric constraint linking $b$ to the spacetime curvature through the metric function $f(r)$:
%%%%
\begin{eqnarray}
b=\sqrt{\frac{r_{0}^{2} } {f\left ( r_{0} \right ) }} \,.
\label{R-bf}
\end{eqnarray}

To quantify the cumulative deflection, we compare the total angular change in curved spacetime to the flat-space baseline $\phi=\pi$. The deflection angle $\bigtriangleup \phi$ is thus expressed as
%%%%
\begin{eqnarray}
\bigtriangleup \phi=2\int_{r_{0}}^{\infty } \left (\frac{r^{4}}{b^{2}}-f\left ( r\right )r^{2}\right )^{-\frac{1}{2} }\mathrm{d}r-\pi\,,
\label{deltaphi}
\end{eqnarray}
where the factor of $2$ accounts for symmetric deflection during approach and recession. To evaluate the integral in the above equation, we implement the dimensionless substitution $ u=\frac{r_{0} }{r} $. In the weak-field regime characterized by $\epsilon \equiv \frac{M}{r_{0} }\ll 1 $, we perform a perturbative expansion of $\bigtriangleup \phi$ to second order in $\epsilon$. This yields the asymptotic expression:
\begin{eqnarray}
\bigtriangleup \phi &=& 2\int_{0}^{1}\left \{ \frac{1}{\sqrt{1-u^{2}}}+\frac{\left ( 1+u+u^{2} \right ) \epsilon }{\left ( 1+u \right ) \sqrt{1-u^{2}} }+\frac{\left [ 3 \left(\frac{1+u+u^{2}}{1+u } \right)^{2}-\left ( 1+u^{2} \right )\zeta^{2}   \right ]\epsilon ^{2} }{2\sqrt{1-u^{2}} }    \right \}\mathrm{d}u \nonumber \\[3mm]
&-&\pi \ +\mathcal{O}\left ( \epsilon ^{3} \right )  \,.  
\label{equ:angle expand}
\end{eqnarray}
Performing the integration in the above equation, we obtain the leading-order quantum-corrected expression for the light deflection angle:
\begin{eqnarray}
\Delta \phi \approx \frac{4M}{r_{0}}\left( 1+\frac{15M\pi}{16r_{0}}-\frac{3\pi M\zeta^{2} }{16r_{0}}-\frac{M}{r_{0}}\right) = \Delta \phi_{\rm{GR}}\left [ 1+\frac{M \left(15\pi - 3\pi \zeta^{2} - 16 M \right)}{16 r_{0}} \right ]   \,,
\label{equ:angle}
\end{eqnarray}
where $\Delta \phi_{\mathrm{GR}}$ represents the standard deflection in GR, with a value of approximately $1.75$ arcsec. We note that the parameter $\epsilon$ has been restored to $\frac{M}{r_{0} }$ in the above expression.

To investigate the detectability of quantum effects and constrain the quantum parameters, we simplify the scenario by defining the closest approach distance $r_0$ as the solar radius (corresponding to light grazing the Sun's limb for detection purposes), while setting $M$ equal to the solar mass. Within the Parameterized Post-Newtonian (PPN) framework, the relativistic gravitational deflection is characterized by the PPN deflection parameter $\gamma$, as illustrated by the expression below:
\begin{eqnarray}
\Delta{\phi}_{\rm{PPN}} \approx \Delta{\phi_{\mathrm{GR}}}\left(\frac{1+\gamma}{2} \right)\,.
\label{eq:ppn}
\end{eqnarray}
Notice that $\gamma$ strictly equals unity ($\gamma=1$) in GR.

Recent advancements in very long baseline interferometry (VLBI) observations of quasar radio waves deflected by the Sun have yielded unprecedented precision in $\gamma$-determinations \cite{Fomalont:2009zg}. By integrating upgraded VLBI observational database and advanced analysis frameworks, the deviation of $\left | \gamma-1 \right |$ has been improved to the order of $10^{-5}$ \cite{lambert2011improved}. By incorporating these results and assuming $\zeta>0$ for quantum gravity effect, we compare Eq. \eqref{equ:angle} with Eq. \eqref{eq:ppn} and directly derive the corresponding bound on $\zeta$ as follows:
\begin{eqnarray}
0< \zeta < 9.12613 .
\end{eqnarray}

%%%%%%%%%%%%%%%%%%%%%%%%%%%%%%%%%%%%%%%%%%
\subsection{Shapiro time delay}

The Shapiro time delay, a fundamental gravitational effect predicted by GR, characterizes the increased propagation time of electromagnetic waves as they traverse the curved spacetime in the vicinity of a massive object. This phenomenon has emerged as one of the cornerstone experimental validations of GR. By measuring the PPN parameters, it is possible to investigate the quantum gravity effect and constrain the quantum corrected parameter.

To investigate this, we analyze the superior conjunction in which the satellite and Earth are positioned on opposite sides of the Sun. The radar signals are emitted from Earth, graze the Sun's gravitational field, and are subsequently reflected by a satellite back to Earth. We begin by deriving the differential equation governing the trajectories of massless particles, expressed in terms of the temporal and radial coordinates $t$ and $r$, through a combination of Eqs. \eqref{Eq:rE} and \eqref{Eq:tE}:
\begin{eqnarray}
\frac{\mathrm{d} t}{\mathrm{d} r} =\pm\frac{1}{f \left ( r \right )\sqrt{ 1-f\left ( r \right )\frac{b^{2}}{r^{2}}  }   }\,,
\end{eqnarray}
where the positive and negative signs correspond to the outgoing and incoming trajectories of the radar waves, respectively.

Then, the propagation time of the electromagnetic signal between the closest approach point $r_{0}$ and either the transmitter location $r_{\mathrm{T}}$ on Earth or the satellite receiver location $r_{\mathrm{R}}$ can be formulated as follows:
\begin{eqnarray}
\Delta t_n = \int_{r_{0}}^{r_n} \frac{1}{f \left ( r \right )\sqrt{ 1-f\left ( r \right )\frac{b^{2}}{r^{2}}  }   }dr\,,
\label{delta tA}
\end{eqnarray}
where $n=\mathrm{T}, \mathrm{R}$. Under weak-field approximation, the propagation time simplifies to:
\begin{eqnarray}
\Delta t_n &\approx& \sqrt{{r_n}^{2}-r_{0}^{2}} +M\left ( \sqrt{\frac{r_n-r_{0}}{r_n+r_{0}} } +2\thinspace \mathrm{arccosh} \left (\frac{r_n}{r_{0}}\right ) \right )   \nonumber \\
&-&\frac{M^{2} \left [\sqrt{\frac{r_n-r_{0}}{r_n+r_{0}} } \left ( \frac{4 r_n + 5r_{0}}{r_n + r_0} \right )+6 \left (-5+\zeta^{2} \right)\mathrm{arcsin}{\left(\frac{\sqrt{1-\frac{r_{0}}{r_n}}}{\sqrt{2} } \right) } \right]}{2r_{0}}\,.
\end{eqnarray}
The leading term $\sqrt{{r_n}^{2}-r_{0}^{2}}$ indicates the travel time of radar signals in flat spacetime, whereas the remaining terms encode the additional relativistic time delay corrected by quantum gravity effects. Consequently, the total round-trip time delay for the radar wave propagation can be formally expressed as:
\begin{eqnarray}
\Delta t_{\mathrm{SC}} &=& 2 \left[ \left(\Delta {t}_{\rm{T}}+ \Delta {t}_{\rm{R}} \right)-\left(\sqrt{{r_{\rm{T}}}^{2}-r_{0}^{2}}+\sqrt{{r_{\rm{R}}}^{2}-r_{0}^{2}} \right) \right] \nonumber \\
&\approx& 4M \left ( 1+\ln{\left ( \frac{4 r_{\mathrm{T}} r_{\mathrm{R}}}{r_{0}^{2}} \right ) } \right )+\frac{M^{2}\left ( 15\pi -8-3\pi \zeta^{2} \right )}{r_{0}}\,.
\label{LQG_sup}
\end{eqnarray} 
For comparison, the parametrized PPN gravitational delay per orbit \cite{ will2018theory, Weinberg:1972kfs} is given by:
\begin{eqnarray}
\Delta t_{\rm{PPN}} \approx 4 M \left[1 + \left(\frac{1+\gamma}{2}\right) \ln\left(\frac{4 r_{\mathrm{T}} r_{\mathrm{R}} }{{r_0}^{2}}\right)\right] .
\label{PPN-formalism}
\end{eqnarray}
Matching the quantum-gravity-corrected result (Eq. \eqref{LQG_sup}) with the PPN framework (Eq. \eqref{PPN-formalism}) yields a direct relation between the PPN parameter $\gamma$ and  the quantum gravity corrected parameter $\zeta$:
\begin{eqnarray}
\gamma-1 = \frac{ M\left ( 15\pi -8-3\pi \zeta^{2}  \right ) }{2 r_0 \ln \left({\frac{4 r_{\mathrm{T}} r_{\mathrm{R}}}{{r_0}^{2}} } \right)}.
\end{eqnarray}

The Cassini solar conjunction experiment, through precision measurements of the Shapiro time delay, currently provides the most stringent observational constraint on the PPN parameter $\gamma$. This yields $\left ( \gamma-1 \right) = \left ( 2.1\pm 2.3 \right)\times 10^{-5}$ relative to the GR prediction \cite{Will:2014kxa, Bertotti:2003rm}. In the actual observations of Cassini's motion, the heliocentric distances of Earth and the spacecraft were precisely determined as $ r_{\mathrm{T}} = 1 \ \mathrm{AU} $ and $ r_{\mathrm{R}} = 8.43 \ \mathrm{AU} $, respectively, while the radio signal attained its closest solar approach with a radius of $r_{0} = 1.6\ R_{\odot}$, where $R_{\odot}$ is the solar radius. Through a systematic comparative analysis, we constrain the quantum-gravity-corrected parameter $\zeta$ to:
\begin{eqnarray}
    0<\zeta< 2.60986.
\end{eqnarray}

The Doppler tracking data from the Cassini spacecraft \cite{1992Relativistic,1993Doppler} provides another approach to constrain the parameter $\zeta$. Unlike direct measurements of the Shapiro time delay, the essence of the Doppler tracking technique lies in measuring the time derivative of Shapiro time delay. Consequently, we obtain the fractional frequency shift of a round-trip radar signal by differentiating Eq. \eqref{LQG_sup} with respect to $t$ \cite{Deng:2015sua,2000Doppler}:
\begin{eqnarray}
\delta \nu = \frac{\Delta \nu}{\nu_{0}} = \frac{\mathrm{d}\Delta t_{\mathrm{SC}} }{\mathrm{d}t} \approx \left [ -\frac{8M}{r_{0}} -\frac{M^{2}\left ( 15\pi -8-3\pi \zeta^{2}  \right ) }{r_{0}^{2}} \right ]\frac{\mathrm{d}r_{0}}{\rm{d}t} \approx \delta \nu_{\rm{GR}}+\delta \nu_{\rm{LQG}} \,,
\end{eqnarray}
where $\Delta \nu \equiv \nu\left( t \right)-\nu_{0}$ quantifies the frequency difference between the Earth-transmitted signal $\nu_0$ and the reflected signal received at time $t$. For spacecraft operating at heliocentric distance that significantly exceed Earth's orbital radius, the time derivative $\mathrm{d}r_{0}/\rm{d}t$ is approximately equivalent to the average orbital velocity of Earth $\nu_\oplus$. We then extract the quantum-corrected frequency shift term, which is given as:
\begin{eqnarray}
\delta \nu_{\rm{LQG}} \approx 
\frac{3M^{2}\pi\zeta^{2} }{r_{0}^{2}}\frac{\mathrm{d}r_{0}}{\rm{d}t} = 
\frac{3M_\odot^{2}\pi\zeta^{2} }{R_\odot^{2}}\frac{256}{729}\nu_\oplus,
\end{eqnarray}
where $M_\odot$ denotes the solar mass. Requiring that this quantum correction $\delta\nu_{\rm{LQG}}$ remain below the experimental sensitivity threshold of $10^{-14}$ yields an upper bound on the quantum-gravity-corrected parameter:
\begin{eqnarray}
0 <\zeta< 2.60\,.
\label{Doppler-Tracking}
\end{eqnarray}
This constraint exhibits remarkable consistency with bounds derived from Shapiro delay measurements, demonstrating complementary validation through independent relativistic observables.

%%%%%%%%%%%%%%%%%%%%%%%%%%%%%%%%%%%%%%%%%%
\subsection{Precession of perihelia}

In this subsection, we leverage the classical GR prediction of Mercury's perihelion precession as a precision testbed to quantify quantum gravity effects. Through systematic analysis of orbital dynamics, we establish constraints on the LQG-corrected parameter $\zeta$. To achieve this, we adopt the standard approximation framework in relativistic celestial mechanics by modeling Mercury as a test particle within the Sun's gravitational field, thereby enabling precise characterization of its geodesic motion.

For timelike geodesics, we have $\eta=1$. It is convenient to adopt the dimensionless inverse radial coordinate $u=\frac{r_0}{r}$ as that in Section \ref{deflection}. Combining Eqs. \eqref{Eq:rE} with \eqref{Eq:phiJ}, the governing differential equation for orbital dynamics takes the following form:
\begin{eqnarray}
\left ( \frac{\mathrm{d}u}{\mathrm{d}\phi} \right )^{2} = \frac{E^{2} {r_0}^{2}}{J^{2}} -f\left ( u \right )\left ( \frac{{r_0}^{2}}{J^{2}}+u^{2}  \right ) \,.
\label{dudphi}
\end{eqnarray}

Given the analytical intractability of the exact solution, we employ the perturbative method to solve the above differential equation \eqref{dudphi}. We begin our perturbative analysis by differentiating Eq. \eqref{dudphi} with respect to the azimuthal angle $\phi$. Implementing the weak-field approximation $\epsilon \equiv M/r_{0}\ll 1$, we derive the following expression for relativistic orbital precession:
\begin{eqnarray}
\frac{\mathrm{d}^{2} u}{\mathrm{d}\phi^{2}}+u-\frac{M^{2}}{J^{2}{\epsilon }} = -\frac{M^{2} u \zeta^{2}}{J^{2}}+ \left(3u^{2} + \frac{6 M^{2} u^{2} \zeta^{2} }{J^{2}} \right)\epsilon + \left ( -\frac{8M^{2}u^{3}\zeta^{2}}{J^{2}}-2u^{3}\zeta^{2} \right ) \epsilon ^{2} +\mathcal{O}\left ( \epsilon ^{3} \right ).
\label{dudphi2}
\end{eqnarray}
The LQG imprint emerges crucially through the $\zeta^2$-dependent terms.

We implement a recursive perturbative scheme by decomposing the orbital function as $u\left(\phi\right)=u_0\left(\phi \right)+u_1\left(\phi \right)$ with $u_0 \left(\phi\right)\ll u_1\left(\phi\right)$, where $u_0\left(\phi\right)$ represents the dominant Newtonian component and $u_1\left(\phi\right)$ encapsulates relativistic-quantum corrections. Truncating at zeroth-order, Eq. \eqref{dudphi2} admits the unperturbed solution as:
\begin{eqnarray}
u_0\left(\phi \right)=\frac{M^2}{J^2 \epsilon} \left(1+e \cos\phi \right) .
\label{sol-u0}
\end{eqnarray}
The above solution corresponds to the Newtonian Keplerian ellipse parametrized by the classical orbital eccentricity $e$.

To systematically quantify relativistic-quantum corrections, we proceed to determine the first-order perturbation $u_1(\phi)$. Implementing the ansatz $u\left(\phi\right)=u_0\left(\phi \right)+u_1\left(\phi \right)$ with $u_0\left(\phi \right)$ given by the Newtonian solution \eqref{sol-u0}, we substitute this decomposition into the precession equation \eqref{dudphi2}. Imposing the boundary conditions $u_1\left(0\right)=0$ and ${\mathrm{d}u_1\left(0\right)}/{\mathrm{d}\phi}=0$ to ensure continuity with the classical trajectory, the perturbative dynamics are governed by:
\begin{eqnarray}
\frac{\mathrm{d}^2 u_1 \left(\phi \right)}{\mathrm{d} \phi^2} + u_1\left(\phi \right) 
= \sum_{i=0}^{3} \mathcal{P}_{i} \cos^{i}\phi ,
\label{equ:III.C-x1-D}
\end{eqnarray}
where the coefficients $\mathcal{P}_{i}$ are given by:
\begin{eqnarray}
&&\mathcal{P}_{0}=\frac{ M^{4}\left [ 3J^{4}-\left ( J^{4}-4J^{2}M^{2}+8M^{4} \right )\zeta ^{2} \right ]     }{J^{8}\epsilon } , \\
&&\mathcal{P}_{1}=\frac{e M^{4} \left [ 6J^{4}-\left ( J^{4}-6J^{2}M^{2}+24M^{4} \right )\zeta ^{2} \right ] }{J^{8}\epsilon }  , \\
&&\mathcal{P}_{2}=\frac{3M^{4}\thinspace e^{2}\left ( J^{4}-8M^{4}\zeta^{2} \right ) }{J^{8}\epsilon } , \\
&&\mathcal{P}_{3}=-\frac{2M^{6}\thinspace e^{3}\zeta^{2}\left ( J^{2}+4M^{2} \right ) }{J^{8}\epsilon }    .  
 \end{eqnarray}
Thus, the perturbed part $u_{1}\left(\phi \right)$ is obtained as:
\begin{eqnarray}
u_1\left(\phi\right) &=& \mathcal{P}_{0} + \frac{\mathcal{P}_{2}}{2}-\mathcal{P}_{0}\cos\phi-\frac{\mathcal{P}_{2}}{3}\cos\phi+\frac{\mathcal{P}_{3}}{32}\cos\phi-\frac{\mathcal{P}_{2}}{6}\cos\left(2\phi\right)-\frac{\mathcal{P}_{3}}{32}\cos\left(3\phi\right)       \nonumber \\[3mm]
&+& \left (\frac{\mathcal{P}_{1}}{2}+ \frac{3\mathcal{P}_{3}}{8} \right )\phi\sin\phi\,.
\label{sol-u1}
\end{eqnarray}

Clearly, the relativistic orbital precession behavior of the test particle depends solely on the sine terms. These non-periodic contributions break the azimuthal symmetry—in their absence, the particle would follow closed Keplerian ellipses characteristic of Newtonian gravity. Furthermore, the cumulative effect would render the periapsis deviation observable. The secular accumulation over orbital cycles manifests as observable periastron advance, providing a critical test of spacetime curvature. Retaining only the dominant the sine terms in Eq. \eqref{sol-u1}, the approximate solution to Eq. \eqref{dudphi2} follows:
\begin{eqnarray}
 u\left(\phi\right) \approx \frac{M r_{0}}{J^2} \left(1 + e \cos\phi \right) + \left( \frac{\mathcal{P}_{1}}{2} + \frac{3 \mathcal{P}_{3}}{8} \right) \phi \sin\phi
\approx \frac{M r_{0}}{J^2} \left \{ 1 + e \cos \left[\left ( 1-\frac{\delta\phi}{2\pi }  \right ) \phi \right] \right \} \,,
\label{sol-u}
\end{eqnarray}
where the cumulative angular precession per orbit
\begin{eqnarray}
\delta \phi\approx \frac{6\pi M^{2}}{J^{2}}\left ( 1-\frac{\zeta ^{2}}{6}  \right )  ,
\label{angular shift}
\end{eqnarray}
quantifies the quantum-gravity modified Einstein precession, recovering the classical result when $\zeta\rightarrow 0$. This analytic expression reveals that the LQG correction ($\zeta^2$ term) suppress the standard relativistic precession rate.
 
To establish a direct connection between orbital geometry and relativistic precession with LQG corrections, we leverage the apsidal extremization scheme derived from the radial extrema in Eq. \eqref{sol-u}.
The minimum (pericenter) $r_{-}$ and maximum (apocenter) $r_{+}$ orbital radii occur at occur at angular positions $\left ( 1-\frac{\delta\phi}{2\pi }  \right ) \phi=0$ and $\left ( 1-\frac{\delta\phi}{2\pi }  \right ) \phi=\pi$ respectively, establishing:
\begin{eqnarray}
r_{-}=\frac{J^{2}}{M\left ( 1+e \right ) } ,
\\
r_{+}=\frac{J^{2}}{M\left ( 1-e \right ) } .
\end{eqnarray}
From these characteristic radii, we can directly derive the semi-major axis $a$ of any elliptical orbit:
\begin{eqnarray}
a = \frac{r_{-} + r_{+}}{2}=\frac{J^2}{M(1-e^2)} .  
\label{equ:III.C-a}
\end{eqnarray}
Then, the perihelion advance per orbital revolution is reformulated as:
\begin{eqnarray}
\Delta \phi=\frac{6\pi M}{a\left ( 1-e^{2} \right ) } \left ( 1-\frac{\zeta ^{2}}{6}  \right )={\Delta \phi_{\mathrm{GR}}}\left ( 1-\frac{\zeta ^{2}}{6}  \right ),
\end{eqnarray}
where
\begin{eqnarray}
{\Delta \phi_{\mathrm{GR}}}=\frac{6\pi M}{a\left ( 1-e^{2} \right ) } .
\end{eqnarray}

Having established the formalism for relativistic perihelion precession with LQG corrections, we we employ high-precision orbital data from the MESSENGER mission to constrain the LQG-corrected parameter $\zeta$. Our analysis of Mercury's anomalous perihelion precession yields a precise measurement of $\Delta{\phi} = \left(42.9799 \pm 0.0009\right)$ arcsec per century \cite{Park:2017zgd}, which imposes the following constraint on $\zeta$:
\begin{eqnarray}
0<\zeta <0.0112089.
\end{eqnarray}

Furthermore, we extend our analysis to Earth-orbiting LAGEOS satellites and the relativistic trajectory of the S2 star orbiting the Galactic Center supermassive BH Sagittarius $\rm{A}^{*}$ (Sgr $\rm{A}^{*}$). For the LAGEOS satellites, the relativistic perigee precession has been precisely constrained using a 13-year laser-ranging dataset \cite{Lucchesi:2010zzb}. The observed anomalous precession rate deviates from GR predictions as:
 \begin{eqnarray}
 \Delta \phi=\Delta \phi_{\mathrm{GR}}\left [ 1+\left ( 0.28\pm 2.14 \right )\times 10^{-3}   \right ] ,
 \end{eqnarray}
results in the following bound for $\zeta$:
\begin{eqnarray}
0 <\zeta < 0.105641 \,.
\end{eqnarray}

In the strong-field regime, the first observational test of relativistic periastron advance was performed using the S2 stellar orbit around the galactic center supermassive BH Sgr $\rm{A}^{*}$ \cite{Do:2019txf, GRAVITY:2018ofz,iorio2024general}. By parameterizing deviations through a post-Newtonian inspired parameter $f_{\rm{SP}}$ ($f_{\rm{SP}}=0$ in Newtonian gravity and $f_{\rm{SP}}=1$ in GR), the GRAVITY collaboration analysis \cite{GRAVITY:2018ofz} establishes a GR-derived periastron shift angle of S2 per orbital period: $\Delta \phi_{\rm{S2}}=\Delta \phi_{\mathrm{GR}}\times f_{\rm{SP}}$, with $\Delta \phi_{\mathrm{GR}}=12.1$ arcmin and $f_{\rm{SP}}=1.1 \pm 0.19$. This results in the following constraint:
 \begin{eqnarray}
0 <\zeta < 0.734847.
 \end{eqnarray}

In summary, within the framework of relativistic perihelion precession, the MESSENGER mission currently provides the most stringent constraint on the LQG parameter $\zeta$, surpassing the constraints from Earth-orbiting LAGEOS satellites and the relativistic trajectory of the S2 star orbiting the Galactic Center supermassive BH Sgr $\rm{A}^{*}$. This can be attributed to the higher experimental accuracy of the MESSENGER mission experiment, as noted in \cite{GRAVITY:2018ofz,Chen:2023bao}. 

Before concluding this section, we note that during the final stages of this work, an independent study by \cite{Xamidov:2025oqx} constrained the parameter $\zeta$. Their analysis used the estimated result of Mercury’s relativistic perihelion precession given by $42.980\pm0.002$ arcsec per century \cite{Benczik:2002tt}, and the GRAVITY data of the pericenter advance of the S2 star orbiting Sgr $\rm{A}^{*}$, reporting $\zeta \leq 0.01869$ (Mercury) and $\zeta \leq 0.73528$ (S2 star), both consistent with our constraints. Additionally, by applying quasiperiodic oscillation (QPO) data from four astrophysical sources (GRO J1655-40,
XTE J1550-564, GRS 1915+105, H1743-322) and Markov Chain Monte Carlo (MCMC) analysis in strong-gravity regimes, they derived $\zeta \leq 2.086$ — a result significantly tighter than previous bounds from BH shadow observations.

In comparison, our work incorporates light deflection measurements from VLBI observations of quasars and Shapiro time-delay data from the Cassini experiment (including Doppler tracking), providing additional independent bounds. This alignment of independently obtained bounds underscores the robustness of current limits on the LQG parameter $\zeta$ and demonstrates consistency across distinct methodologies.

%%%%%%%%%%%%%%%%%%%%%%%%%%%%%%%%%%%%%%%%%%
\section{Conclusion}\label{conclusion} 

%%%----------------------------------------------------------------------------------------%%
\begin{table}
	\centering
	%\begin{tabular}{cccc}
	%  \setlength{\tabnotewidth}{1.0\columnwidth}
	%  \tablecols{4}
	%   \setlength{\tabcolsep}{2.8\tabcolsep}
	%\caption{C\lowercase{onstituents of the universe and  their behaviour: 
			%density evolution $\rho(a)$, scale factor $a(t)$}, H\lowercase{ubble parameter} $H(\lowercase{t})$.}
	\caption{\centering Summary of observational constraints on the LQG-corrected parameter $\zeta$. Datasets are drawn from solar system and galactic center observations, with corresponding relative uncertainties.}
	\vspace{0.5em}
	\begin{tabular}{c|c|c|c}
		\toprule[1pt]
		\toprule[0.5pt]
		\midrule
		\quad {\rm Experiments/Observations \ } \quad & \ Relative uncertainties \ & \ $ \zeta $  \ & \quad Datasets \\
		\midrule
		\toprule[1pt]
		\midrule
		Light deflection &   $0.012 \%$ & \quad $ 9.12613 \quad$ &  \   VLBI observation of quasars	\\	
		\midrule
		\toprule[0.1pt]
		\midrule
		\multirow{2}{*}{Shapiro time delay} &  $0.0023\%$ &  \quad  $ 2.60986 \quad$  & \  Cassini experiment	\\	
		&  /  & \quad  $  2.60  \quad$  & \ Doppler tracking of Cassini 	\\
		\midrule
		\toprule[0.1pt]
		\midrule
		\multirow{3}{*}{Perihelion advance} &  $ 0.0021\%$ & \quad $ 0.0112089 \quad$ &\ MESSENGER mission  \\
		&   $0.214\%$ & \quad $ 0.105641 \quad$  &\ LAGEOS satellites  \\
		&   $17.27\%$ & \quad $ 0.734847 \quad$  & \  Observation of S2 around Sgr $\rm{A}^{*}$ \\
		\midrule
		\toprule[0.1pt]
		\midrule
		
		\bottomrule[1pt]
	\end{tabular}
	\label{tab:summary}
	%\end{ruledtabular}
\end{table}
%%%----------------------------------------------------------------------------------------%%

Classical tests of GR — including light deflection, Shapiro time delay, and perihelion precession — provide fundamental laboratories for probing gravitational theories ranging from GR itself to alternative classical gravities and quantum gravity candidates. In this paper, we investigate quantum gravity effects through an effective LQG-BH model, leveraging precision measurements from these classical GR experiments. Additionally, we also extend our analysis to Earth-orbiting LAGEOS satellites and the relativistic trajectory of the S2 star orbiting the Galactic Center supermassive black hole Sgr $\rm{A}^{*}$. The constraint results are summarized in Table \ref{tab:summary}.

Theoretical calculations reveal that the inclusion of ${\zeta}^2$ terms induces deviations from GR predictions, resulting in a lagged manifestation of relativistic phenomena. As summarized in Table \ref{tab:summary}, the tightest constraint on $\zeta$ arises from the MESSENGER mission data of the Mercury periasis shift, yielding $0<\zeta<0.0112089$, whereas the second most stringent constraint is produced by the LAGEOS satellites, at the level of $10^{-1}$. To further constrain $\zeta$, we employ the strong gravitational field observations of the S2 star orbit around Sgr $\rm{A}^{*}$, deriving an upper bound $\zeta \lesssim 10^{-1}$, yielding a tighter constraint than those obtained from EHT data of BH shadow radius \cite{Konoplya:2024lch, Heidari:2024bkm, Wang:2024iwt, Shu:2024tut, Chen:2025ifv}.

It is valuable to contextualize theses observational bounds with theoretical expectations. Fixing the BI parameter at $\gamma = 0.2375$ \cite{Domagala:2004jt}, Eq. \eqref{zeta_M} gives the characteristic scaling:
\begin{equation}
	\zeta_{\mathrm{theo}}^{\odot} \sim \frac{\gamma \thinspace\mathcal{O}\left(A^{\frac{1}{2}}\right)}{{M}_{\odot}} \approx 10^{-39} \,,
\end{equation}
where we substitute $A = 4\sqrt{3}\pi\gamma\ell_P^2$ with $\ell_P \sim 10^{-35}$ m and solar mass ${M}_{\odot} \sim 10^{30}$ kg. This reveals the critical mass dependence of quantum gravity effects:
	\begin{itemize}
		\item Theoretical $\zeta \propto 1/M$ explains why solar-system constraints ($\zeta < 0.01$) remain $\sim 37$ orders above $\zeta_{\mathrm{theo}}^{\odot}$.
		\item Quantum gravity effects become significant when $M \lesssim \gamma\sqrt{A}/\zeta_{\mathrm{obs}} \sim (10^{-6}-10^{-8})$ kg (Planck-mass scales).
	\end{itemize}
While current solar-system tests cannot probe fundamental $\zeta$ values, our constraints establish empirical upper bounds for phenomenological LQG models. Future work will prioritize strong-field regimes, particularly GW signatures from BH mergers.

Although current solar system experiments lack the sensitivity to detect LQG signatures, upcoming space-based missions will progressively tighten constraints on quantum-corrected parameters. For instance, the optical interferometry-based Gaia orbital telescope \cite{Gaia:2016zol}, with its superb light deflection measurements, is projected to achieve the PPN parameter $\gamma$ with an accuracy in the range of $10^{-5} \sim 10^{-7}$. The Mercury Orbiter Radio-science Experiment (MORE) of BepiColombo will exploit solar conjunctions to measure radio gravitational delays, with predicted $\gamma$ constraints reaching $10^{-6}$ level \cite{Will:2018mcj, Genova:2021rlo}. Furthermore, complementary efforts by Jupiter orbiting missions such as Juno and JUICE, will further refine $\gamma$ estimates via spacecraft trajectory tracking, with the latter targeting precision at the $10^{-7}$ level \cite{DeMarchi:2019lei}. Synthesizing these multi-mission datasets could improve current $\zeta$ bounds by 1--2 orders of magnitude. Meanwhile, next-generation GW detectors (e.g., LISA \cite{LISA:2022yao}, Einstein Telescope \cite{ET:2019dnz}) may directly probe such signatures and impose even stronger constraints.

Future investigations could extend this research program along the following promising avenues. First, the methodology developed herein may be extended to analyze quantum corrections in gravitational time delay effects generated by spinning oblate masses as \cite{Battista:2017xlm}. Second, a particularly relevant application would involve re-examining the Newtonian Lagrangian points within the Earth-Moon system through the lens of quantum gravity phenomenology, potentially revealing observable signatures in celestial mechanics \cite{Battista:2014oba,Battista:2014ija,Battista:2015zta,Battista:2015wwa,Tartaglia:2018bjc,Battista:2020qqp}. In addition, the emerging framework of LQG could be rigorously tested through detailed simulations of extreme mass-ratio inspirals, where recent theoretical advances \cite{Fu:2024cfk,Zi:2024jla} suggest new observational windows for probing quantum spacetime structure.

\acknowledgments

We are especially grateful to Prof. Rui-Hong Yue for helpful discussions and suggestions. This work is supported by the Natural Science Foundation of China under Grants No. 12375055.

%%%%%%%%%%%%%%%%%%%%%%%%%%%%%%%%%%%%%%%%%%

\bibliographystyle{utphys}
\bibliography{ref}

\providecommand{\href}[2]{#2}\begingroup\raggedright\begin{thebibliography}{10}

\bibitem{Will:2014kxa}
C.~M. Will, ``{The Confrontation between General Relativity and Experiment},''
  \href{http://dx.doi.org/10.12942/lrr-2014-4}{{\em Living Rev. Rel.}
  {\bfseries 17} (2014) 4}, \href{http://arxiv.org/abs/1403.7377}{{\ttfamily
  arXiv:1403.7377 [gr-qc]}}.

\bibitem{Park:2017zgd}
R.~S. Park, W.~M. Folkner, A.~S. Konopliv, J.~G. Williams, D.~E. Smith, and
  M.~T. Zuber, ``{Precession of Mercury\textquoteright{}s Perihelion from
  Ranging to the MESSENGER Spacecraft},''
  \href{http://dx.doi.org/10.3847/1538-3881/aa5be2}{{\em Astron. J.} {\bfseries
  153} no.~3, (2017) 121}.

\bibitem{Fomalont:2009zg}
E.~Fomalont, S.~Kopeikin, G.~Lanyi, and J.~Benson, ``{Progress in Measurements
  of the Gravitational Bending of Radio Waves Using the VLBA},''
  \href{http://dx.doi.org/10.1088/0004-637X/699/2/1395}{{\em Astrophys. J.}
  {\bfseries 699} (2009) 1395--1402},
  \href{http://arxiv.org/abs/0904.3992}{{\ttfamily arXiv:0904.3992
  [astro-ph.CO]}}.

\bibitem{Bertotti:2003rm}
B.~Bertotti, L.~Iess, and P.~Tortora, ``{A test of general relativity using
  radio links with the Cassini spacecraft},''
  \href{http://dx.doi.org/10.1038/nature01997}{{\em Nature} {\bfseries 425}
  (2003) 374--376}.

\bibitem{Stairs:2003eg}
I.~H. Stairs, ``{Testing general relativity with pulsar timing},''
  \href{http://dx.doi.org/10.12942/lrr-2003-5}{{\em Living Rev. Rel.}
  {\bfseries 6} (2003) 5},
  \href{http://arxiv.org/abs/astro-ph/0307536}{{\ttfamily
  arXiv:astro-ph/0307536}}.

\bibitem{EventHorizonTelescope:2019dse}
{\bfseries Event Horizon Telescope} Collaboration, K.~Akiyama {\em et~al.},
  ``{First M87 Event Horizon Telescope Results. I. The Shadow of the
  Supermassive Black Hole},''
  \href{http://dx.doi.org/10.3847/2041-8213/ab0ec7}{{\em Astrophys. J. Lett.}
  {\bfseries 875} (2019) L1}, \href{http://arxiv.org/abs/1906.11238}{{\ttfamily
  arXiv:1906.11238 [astro-ph.GA]}}.

\bibitem{EventHorizonTelescope:2020qrl}
{\bfseries Event Horizon Telescope} Collaboration, D.~Psaltis {\em et~al.},
  ``{Gravitational Test Beyond the First Post-Newtonian Order with the Shadow
  of the M87 Black Hole},''
  \href{http://dx.doi.org/10.1103/PhysRevLett.125.141104}{{\em Phys. Rev.
  Lett.} {\bfseries 125} no.~14, (2020) 141104},
  \href{http://arxiv.org/abs/2010.01055}{{\ttfamily arXiv:2010.01055 [gr-qc]}}.

\bibitem{LIGOScientific:2016aoc}
{\bfseries LIGO Scientific, Virgo} Collaboration, B.~P. Abbott {\em et~al.},
  ``{Observation of Gravitational Waves from a Binary Black Hole Merger},''
  \href{http://dx.doi.org/10.1103/PhysRevLett.116.061102}{{\em Phys. Rev.
  Lett.} {\bfseries 116} no.~6, (2016) 061102},
  \href{http://arxiv.org/abs/1602.03837}{{\ttfamily arXiv:1602.03837 [gr-qc]}}.

\bibitem{Borde:1993xh}
A.~Borde and A.~Vilenkin, ``{Eternal inflation and the initial singularity},''
  \href{http://dx.doi.org/10.1103/PhysRevLett.72.3305}{{\em Phys. Rev. Lett.}
  {\bfseries 72} (1994) 3305--3309},
  \href{http://arxiv.org/abs/gr-qc/9312022}{{\ttfamily arXiv:gr-qc/9312022}}.

\bibitem{Hawking:1973uf}
S.~W. Hawking and G.~F.~R. Ellis,
  \href{http://dx.doi.org/10.1017/9781009253161}{{\em {The Large Scale
  Structure of Space-Time}}}.
\newblock Cambridge Monographs on Mathematical Physics. Cambridge University
  Press, 2, 2023.

\bibitem{Adler:2010wf}
R.~J. Adler, ``{Six easy roads to the Planck scale},''
  \href{http://dx.doi.org/10.1119/1.3439650}{{\em Am. J. Phys.} {\bfseries 78}
  (2010) 925--932}, \href{http://arxiv.org/abs/1001.1205}{{\ttfamily
  arXiv:1001.1205 [gr-qc]}}.

\bibitem{Ng:2003jk}
Y.~J. Ng, ``{Selected topics in Planck scale physics},''
  \href{http://dx.doi.org/10.1142/S0217732303010934}{{\em Mod. Phys. Lett. A}
  {\bfseries 18} (2003) 1073--1098},
  \href{http://arxiv.org/abs/gr-qc/0305019}{{\ttfamily arXiv:gr-qc/0305019}}.

\bibitem{Thiemann:2007pyv}
T.~Thiemann, \href{http://dx.doi.org/10.1017/CBO9780511755682}{{\em {Modern
  Canonical Quantum General Relativity}}}.
\newblock Cambridge Monographs on Mathematical Physics. Cambridge University
  Press, 2007.

\bibitem{Smolin:2006pa}
L.~Smolin, ``{Generic predictions of quantum theories of gravity},''
  \href{http://arxiv.org/abs/hep-th/0605052}{{\ttfamily arXiv:hep-th/0605052}}.

\bibitem{Han:2005km}
M.~Han, W.~Huang, and Y.~Ma, ``{Fundamental structure of loop quantum
  gravity},'' \href{http://dx.doi.org/10.1142/S0218271807010894}{{\em Int. J.
  Mod. Phys. D} {\bfseries 16} (2007) 1397--1474},
  \href{http://arxiv.org/abs/gr-qc/0509064}{{\ttfamily arXiv:gr-qc/0509064}}.

\bibitem{Bojowald:2001xe}
M.~Bojowald, ``{Absence of singularity in loop quantum cosmology},''
  \href{http://dx.doi.org/10.1103/PhysRevLett.86.5227}{{\em Phys. Rev. Lett.}
  {\bfseries 86} (2001) 5227--5230},
  \href{http://arxiv.org/abs/gr-qc/0102069}{{\ttfamily arXiv:gr-qc/0102069}}.

\bibitem{Ashtekar:2006rx}
A.~Ashtekar, T.~Pawlowski, and P.~Singh, ``{Quantum nature of the big bang},''
  \href{http://dx.doi.org/10.1103/PhysRevLett.96.141301}{{\em Phys. Rev. Lett.}
  {\bfseries 96} (2006) 141301},
  \href{http://arxiv.org/abs/gr-qc/0602086}{{\ttfamily arXiv:gr-qc/0602086}}.

\bibitem{Ashtekar:2006uz}
A.~Ashtekar, T.~Pawlowski, and P.~Singh, ``{Quantum Nature of the Big Bang: An
  Analytical and Numerical Investigation. I.},''
  \href{http://dx.doi.org/10.1103/PhysRevD.73.124038}{{\em Phys. Rev. D}
  {\bfseries 73} (2006) 124038},
  \href{http://arxiv.org/abs/gr-qc/0604013}{{\ttfamily arXiv:gr-qc/0604013}}.

\bibitem{Ashtekar:2006wn}
A.~Ashtekar, T.~Pawlowski, and P.~Singh, ``{Quantum Nature of the Big Bang:
  Improved dynamics},''
  \href{http://dx.doi.org/10.1103/PhysRevD.74.084003}{{\em Phys. Rev. D}
  {\bfseries 74} (2006) 084003},
  \href{http://arxiv.org/abs/gr-qc/0607039}{{\ttfamily arXiv:gr-qc/0607039}}.

\bibitem{Ashtekar:2003hd}
A.~Ashtekar, M.~Bojowald, and J.~Lewandowski, ``{Mathematical structure of loop
  quantum cosmology},''
  \href{http://dx.doi.org/10.4310/ATMP.2003.v7.n2.a2}{{\em Adv. Theor. Math.
  Phys.} {\bfseries 7} no.~2, (2003) 233--268},
  \href{http://arxiv.org/abs/gr-qc/0304074}{{\ttfamily arXiv:gr-qc/0304074}}.

\bibitem{Bojowald:2005epg}
M.~Bojowald, ``{Loop quantum cosmology},''
  \href{http://dx.doi.org/10.12942/lrr-2005-11}{{\em Living Rev. Rel.}
  {\bfseries 8} (2005) 11},
  \href{http://arxiv.org/abs/gr-qc/0601085}{{\ttfamily arXiv:gr-qc/0601085}}.

\bibitem{Ashtekar:2011ni}
A.~Ashtekar and P.~Singh, ``{Loop Quantum Cosmology: A Status Report},''
  \href{http://dx.doi.org/10.1088/0264-9381/28/21/213001}{{\em Class. Quant.
  Grav.} {\bfseries 28} (2011) 213001},
  \href{http://arxiv.org/abs/1108.0893}{{\ttfamily arXiv:1108.0893 [gr-qc]}}.

\bibitem{Ashtekar:2015dja}
A.~Ashtekar and A.~Barrau, ``{Loop quantum cosmology: From pre-inflationary
  dynamics to observations},''
  \href{http://dx.doi.org/10.1088/0264-9381/32/23/234001}{{\em Class. Quant.
  Grav.} {\bfseries 32} no.~23, (2015) 234001},
  \href{http://arxiv.org/abs/1504.07559}{{\ttfamily arXiv:1504.07559 [gr-qc]}}.

\bibitem{Chiou:2008nm}
D.-W. Chiou, ``{Phenomenological loop quantum geometry of the Schwarzschild
  black hole},'' \href{http://dx.doi.org/10.1103/PhysRevD.78.064040}{{\em Phys.
  Rev. D} {\bfseries 78} (2008) 064040},
  \href{http://arxiv.org/abs/0807.0665}{{\ttfamily arXiv:0807.0665 [gr-qc]}}.

\bibitem{Chiou:2008eg}
D.-W. Chiou, ``{Phenomenological dynamics of loop quantum cosmology in
  Kantowski-Sachs spacetime},''
  \href{http://dx.doi.org/10.1103/PhysRevD.78.044019}{{\em Phys. Rev. D}
  {\bfseries 78} (2008) 044019},
  \href{http://arxiv.org/abs/0803.3659}{{\ttfamily arXiv:0803.3659 [gr-qc]}}.

\bibitem{Boehmer:2007ket}
C.~G. Boehmer and K.~Vandersloot, ``{Loop Quantum Dynamics of the Schwarzschild
  Interior},'' \href{http://dx.doi.org/10.1103/PhysRevD.76.104030}{{\em Phys.
  Rev. D} {\bfseries 76} (2007) 104030},
  \href{http://arxiv.org/abs/0709.2129}{{\ttfamily arXiv:0709.2129 [gr-qc]}}.

\bibitem{Perez:2017cmj}
A.~Perez, ``{Black Holes in Loop Quantum Gravity},''
  \href{http://dx.doi.org/10.1088/1361-6633/aa7e14}{{\em Rept. Prog. Phys.}
  {\bfseries 80} no.~12, (2017) 126901},
  \href{http://arxiv.org/abs/1703.09149}{{\ttfamily arXiv:1703.09149 [gr-qc]}}.

\bibitem{Modesto:2005zm}
L.~Modesto, ``{Loop quantum black hole},''
  \href{http://dx.doi.org/10.1088/0264-9381/23/18/006}{{\em Class. Quant.
  Grav.} {\bfseries 23} (2006) 5587--5602},
  \href{http://arxiv.org/abs/gr-qc/0509078}{{\ttfamily arXiv:gr-qc/0509078}}.

\bibitem{Modesto:2008im}
L.~Modesto, ``{Semiclassical loop quantum black hole},''
  \href{http://dx.doi.org/10.1007/s10773-010-0346-x}{{\em Int. J. Theor. Phys.}
  {\bfseries 49} (2010) 1649--1683},
  \href{http://arxiv.org/abs/0811.2196}{{\ttfamily arXiv:0811.2196 [gr-qc]}}.

\bibitem{Ashtekar:2005qt}
A.~Ashtekar and M.~Bojowald, ``{Quantum geometry and the Schwarzschild
  singularity},'' \href{http://dx.doi.org/10.1088/0264-9381/23/2/008}{{\em
  Class. Quant. Grav.} {\bfseries 23} (2006) 391--411},
  \href{http://arxiv.org/abs/gr-qc/0509075}{{\ttfamily arXiv:gr-qc/0509075}}.

\bibitem{Zhang:2023yps}
X.~Zhang, ``{Loop Quantum Black Hole},''
  \href{http://dx.doi.org/10.3390/universe9070313}{{\em Universe} {\bfseries 9}
  no.~7, (2023) 313}, \href{http://arxiv.org/abs/2308.10184}{{\ttfamily
  arXiv:2308.10184 [gr-qc]}}.

\bibitem{Zhang:2023okw}
C.~Zhang, Y.~Ma, and J.~Yang, ``{Black hole image encoding quantum gravity
  information},'' \href{http://dx.doi.org/10.1103/PhysRevD.108.104004}{{\em
  Phys. Rev. D} {\bfseries 108} no.~10, (2023) 104004},
  \href{http://arxiv.org/abs/2302.02800}{{\ttfamily arXiv:2302.02800 [gr-qc]}}.

\bibitem{Zhang:2024ney}
C.~Zhang, J.~Lewandowski, Y.~Ma, and J.~Yang, ``{Black holes and covariance in
  effective quantum gravity: A solution without Cauchy horizons},''
  \href{http://arxiv.org/abs/2412.02487}{{\ttfamily arXiv:2412.02487 [gr-qc]}}.

\bibitem{Ashtekar:2020gec}
A.~Ashtekar, B.~Gupt, D.~Jeong, and V.~Sreenath, ``{Alleviating the Tension in
  the Cosmic Microwave Background using Planck-Scale Physics},''
  \href{http://dx.doi.org/10.1103/PhysRevLett.125.051302}{{\em Phys. Rev.
  Lett.} {\bfseries 125} no.~5, (2020) 051302},
  \href{http://arxiv.org/abs/2001.11689}{{\ttfamily arXiv:2001.11689
  [astro-ph.CO]}}.

\bibitem{Ashtekar:2021izi}
A.~Ashtekar, B.~Gupt, and V.~Sreenath, ``{Cosmic Tango Between the Very Small
  and the Very Large: Addressing CMB Anomalies Through Loop Quantum
  Cosmology},'' \href{http://dx.doi.org/10.3389/fspas.2021.685288}{{\em Front.
  Astron. Space Sci.} {\bfseries 8} (2021) 76},
  \href{http://arxiv.org/abs/2103.14568}{{\ttfamily arXiv:2103.14568 [gr-qc]}}.

\bibitem{Agullo:2020fbw}
I.~Agullo, D.~Kranas, and V.~Sreenath, ``{Anomalies in the CMB from a cosmic
  bounce},'' \href{http://dx.doi.org/10.1007/s10714-020-02778-9}{{\em Gen. Rel.
  Grav.} {\bfseries 53} no.~2, (2021) 17},
  \href{http://arxiv.org/abs/2005.01796}{{\ttfamily arXiv:2005.01796
  [astro-ph.CO]}}.

\bibitem{Agullo:2021oqk}
I.~Agullo, D.~Kranas, and V.~Sreenath, ``{Anomalies in the Cosmic Microwave
  Background and Their Non-Gaussian Origin in Loop Quantum Cosmology},''
  \href{http://dx.doi.org/10.3389/fspas.2021.703845}{{\em Front. Astron. Space
  Sci.} {\bfseries 8} (2021) 703845},
  \href{http://arxiv.org/abs/2105.12993}{{\ttfamily arXiv:2105.12993 [gr-qc]}}.

\bibitem{Zhang:2007bi}
X.~Zhang and Y.~Ling, ``{Inflationary universe in loop quantum cosmology},''
  \href{http://dx.doi.org/10.1088/1475-7516/2007/08/012}{{\em JCAP} {\bfseries
  08} (2007) 012}, \href{http://arxiv.org/abs/0705.2656}{{\ttfamily
  arXiv:0705.2656 [gr-qc]}}.

\bibitem{Li:2011ac}
L.-F. Li, R.-G. Cai, Z.-K. Guo, and B.~Hu, ``{Non-Gaussian features from the
  inverse volume corrections in loop quantum cosmology},''
  \href{http://dx.doi.org/10.1103/PhysRevD.86.044020}{{\em Phys. Rev. D}
  {\bfseries 86} (2012) 044020},
  \href{http://arxiv.org/abs/1112.2785}{{\ttfamily arXiv:1112.2785 [gr-qc]}}.

\bibitem{deBlas:2016puz}
D.~M. de~Blas and J.~Olmedo, ``{Primordial power spectra for scalar
  perturbations in loop quantum cosmology},''
  \href{http://dx.doi.org/10.1088/1475-7516/2016/06/029}{{\em JCAP} {\bfseries
  06} (2016) 029}, \href{http://arxiv.org/abs/1601.01716}{{\ttfamily
  arXiv:1601.01716 [gr-qc]}}.

\bibitem{Martin-Benito:2021szh}
M.~Mart{\'\i}n-Benito, R.~B. Neves, and J.~Olmedo, ``{States of Low Energy in
  bouncing inflationary scenarios in Loop Quantum Cosmology},''
  \href{http://dx.doi.org/10.1103/PhysRevD.103.123524}{{\em Phys. Rev. D}
  {\bfseries 103} (2021) 123524},
  \href{http://arxiv.org/abs/2104.03035}{{\ttfamily arXiv:2104.03035 [gr-qc]}}.

\bibitem{LIGOScientific:2016lio}
{\bfseries LIGO Scientific, Virgo} Collaboration, B.~P. Abbott {\em et~al.},
  ``{Tests of general relativity with GW150914},''
  \href{http://dx.doi.org/10.1103/PhysRevLett.116.221101}{{\em Phys. Rev.
  Lett.} {\bfseries 116} no.~22, (2016) 221101},
  \href{http://arxiv.org/abs/1602.03841}{{\ttfamily arXiv:1602.03841 [gr-qc]}}.
  [Erratum: Phys. Rev. Lett. \textbf{121} (2018) 129902 ].

\bibitem{LIGOScientific:2016sjg}
{\bfseries LIGO Scientific, Virgo} Collaboration, B.~P. Abbott {\em et~al.},
  ``{GW151226: Observation of Gravitational Waves from a 22-Solar-Mass Binary
  Black Hole Coalescence},''
  \href{http://dx.doi.org/10.1103/PhysRevLett.116.241103}{{\em Phys. Rev.
  Lett.} {\bfseries 116} no.~24, (2016) 241103},
  \href{http://arxiv.org/abs/1606.04855}{{\ttfamily arXiv:1606.04855 [gr-qc]}}.

\bibitem{EventHorizonTelescope:2019ths}
{\bfseries Event Horizon Telescope} Collaboration, K.~Akiyama {\em et~al.},
  ``{First M87 Event Horizon Telescope Results. IV. Imaging the Central
  Supermassive Black Hole},''
  \href{http://dx.doi.org/10.3847/2041-8213/ab0e85}{{\em Astrophys. J. Lett.}
  {\bfseries 875} no.~1, (2019) L4},
  \href{http://arxiv.org/abs/1906.11241}{{\ttfamily arXiv:1906.11241
  [astro-ph.GA]}}.

\bibitem{EventHorizonTelescope:2022xnr}
{\bfseries Event Horizon Telescope} Collaboration, K.~Akiyama {\em et~al.},
  ``{First Sagittarius A* Event Horizon Telescope Results. I. The Shadow of the
  Supermassive Black Hole in the Center of the Milky Way},''
  \href{http://dx.doi.org/10.3847/2041-8213/ac6674}{{\em Astrophys. J. Lett.}
  {\bfseries 930} no.~2, (2022) L12}.

\bibitem{EventHorizonTelescope:2022xqj}
{\bfseries Event Horizon Telescope} Collaboration, K.~Akiyama {\em et~al.},
  ``{First Sagittarius A* Event Horizon Telescope Results. VI. Testing the
  Black Hole Metric},'' \href{http://dx.doi.org/10.3847/2041-8213/ac6756}{{\em
  Astrophys. J. Lett.} {\bfseries 930} no.~2, (2022) L17}.

\bibitem{Konoplya:2024lch}
R.~A. Konoplya and O.~S. Stashko, ``{Probing the Effective Quantum Gravity via
  Quasinormal Modes and Shadows of Black Holes},''
  \href{http://arxiv.org/abs/2408.02578}{{\ttfamily arXiv:2408.02578 [gr-qc]}}.

\bibitem{Skvortsova:2024msa}
M.~Skvortsova, ``{Quantum corrected black holes: testing the correspondence
  between grey-body factors and quasinormal modes},''
  \href{http://arxiv.org/abs/2411.06007}{{\ttfamily arXiv:2411.06007 [gr-qc]}}.

\bibitem{Zhang:2024svj}
C.~Zhang and A.~Wang, ``{Quasi-normal modes of loop quantum black holes formed
  from gravitational collapse},''
  \href{http://dx.doi.org/10.1088/1475-7516/2024/10/070}{{\em JCAP} {\bfseries
  10} (2024) 070}, \href{http://arxiv.org/abs/2407.19654}{{\ttfamily
  arXiv:2407.19654 [gr-qc]}}.

\bibitem{Fu:2023drp}
G.~Fu, D.~Zhang, P.~Liu, X.-M. Kuang, and J.-P. Wu, ``{Peculiar properties in
  quasi-normal spectra from loop quantum gravity effect},''
  \href{http://arxiv.org/abs/2301.08421}{{\ttfamily arXiv:2301.08421 [gr-qc]}}.

\bibitem{Gong:2023ghh}
H.~Gong, S.~Li, D.~Zhang, G.~Fu, and J.-P. Wu, ``{Quasinormal modes of
  quantum-corrected black holes},''
  \href{http://dx.doi.org/10.1103/PhysRevD.110.044040}{{\em Phys. Rev. D}
  {\bfseries 110} no.~4, (2024) 044040},
  \href{http://arxiv.org/abs/2312.17639}{{\ttfamily arXiv:2312.17639 [gr-qc]}}.

\bibitem{Zhu:2024wic}
L.-G. Zhu, G.~Fu, S.~Li, D.~Zhang, and J.-P. Wu, ``{Quasinormal modes of a
  charged loop quantum black hole},''
  \href{http://dx.doi.org/10.1103/PhysRevD.111.104008}{{\em Phys. Rev. D}
  {\bfseries 111} no.~10, (2025) 104008},
  \href{http://arxiv.org/abs/2410.00543}{{\ttfamily arXiv:2410.00543 [gr-qc]}}.

\bibitem{Shao:2023qlt}
C.-Y. Shao, C.~Zhang, W.~Zhang, and C.-G. Shao, ``{Scalar fields around a loop
  quantum gravity black hole in de Sitter spacetime: Quasinormal modes,
  late-time tails and strong cosmic censorship},''
  \href{http://dx.doi.org/10.1103/PhysRevD.109.064012}{{\em Phys. Rev. D}
  {\bfseries 109} no.~6, (2024) 064012},
  \href{http://arxiv.org/abs/2309.04962}{{\ttfamily arXiv:2309.04962 [gr-qc]}}.

\bibitem{Liu:2024soc}
W.~Liu, D.~Wu, and J.~Wang, ``{Light rings and shadows of static black holes in
  effective quantum gravity},''
  \href{http://dx.doi.org/10.1016/j.physletb.2024.139052}{{\em Phys. Lett. B}
  {\bfseries 858} (2024) 139052},
  \href{http://arxiv.org/abs/2408.05569}{{\ttfamily arXiv:2408.05569 [gr-qc]}}.

\bibitem{Liu:2024wal}
H.~Liu, M.-Y. Lai, X.-Y. Pan, H.~Huang, and D.-C. Zou, ``{Gravitational lensing
  effect of black holes in effective quantum gravity},''
  \href{http://dx.doi.org/10.1103/PhysRevD.110.104039}{{\em Phys. Rev. D}
  {\bfseries 110} no.~10, (2024) 104039},
  \href{http://arxiv.org/abs/2408.11603}{{\ttfamily arXiv:2408.11603 [gr-qc]}}.

\bibitem{Xamidov:2025oqx}
T.~Xamidov, S.~Shaymatov, B.~Ahmedov, and T.~Zhu, ``{Probing quantum corrected
  black hole through astrophysical tests with the orbit of S2 star and
  quasiperiodic oscillations},''
  \href{http://arxiv.org/abs/2503.06750}{{\ttfamily arXiv:2503.06750 [gr-qc]}}.

\bibitem{Du:2024ujg}
Y.~Du, Y.~Liu, and X.~Zhang, ``{Spinning Particle Dynamics and ISCO in
  Covariant Loop Quantum Gravity},''
  \href{http://arxiv.org/abs/2411.13316}{{\ttfamily arXiv:2411.13316 [gr-qc]}}.

\bibitem{Shu:2024tut}
Y.-H. Shu and J.-H. Huang, ``{Circular orbits and thin accretion disk around a
  quantum corrected black hole},''
  \href{http://dx.doi.org/10.1016/j.physletb.2025.139411}{{\em Phys. Lett. B}
  {\bfseries 864} (2025) 139411},
  \href{http://arxiv.org/abs/2412.05670}{{\ttfamily arXiv:2412.05670 [gr-qc]}}.

\bibitem{Chen:2025ifv}
J.~Chen and J.~Yang, ``{Shadows and optical appearance of quantum-corrected
  black holes illuminated by static thin accretions},''
  \href{http://arxiv.org/abs/2503.06215}{{\ttfamily arXiv:2503.06215 [gr-qc]}}.

\bibitem{Yang:2022btw}
J.~Yang, C.~Zhang, and Y.~Ma, ``{Shadow and stability of quantum-corrected
  black holes},'' \href{http://dx.doi.org/10.1140/epjc/s10052-023-11800-8}{{\em
  Eur. Phys. J. C} {\bfseries 83} no.~7, (2023) 619},
  \href{http://arxiv.org/abs/2211.04263}{{\ttfamily arXiv:2211.04263 [gr-qc]}}.

\bibitem{Tu:2023xab}
Z.-Y. Tu, T.~Zhu, and A.~Wang, ``{Periodic orbits and their gravitational wave
  radiations in a polymer black hole in loop quantum gravity},''
  \href{http://dx.doi.org/10.1103/PhysRevD.108.024035}{{\em Phys. Rev. D}
  {\bfseries 108} no.~2, (2023) 024035},
  \href{http://arxiv.org/abs/2304.14160}{{\ttfamily arXiv:2304.14160 [gr-qc]}}.

\bibitem{Fu:2024cfk}
G.~Fu, Y.~Liu, B.~Wang, J.-P. Wu, and C.~Zhang, ``{Probing quantum gravity
  effects with eccentric extreme mass-ratio inspirals},''
  \href{http://dx.doi.org/10.1103/PhysRevD.111.084066}{{\em Phys. Rev. D}
  {\bfseries 111} no.~8, (2025) 084066},
  \href{http://arxiv.org/abs/2409.08138}{{\ttfamily arXiv:2409.08138 [gr-qc]}}.

\bibitem{Zi:2024jla}
T.~Zi and S.~Kumar, ``{Eccentric extreme mass-ratio inspirals: A gateway to
  probe quantum gravity effects},''
  \href{http://arxiv.org/abs/2409.17765}{{\ttfamily arXiv:2409.17765 [gr-qc]}}.

\bibitem{Zhu:2020tcf}
T.~Zhu and A.~Wang, ``{Observational tests of the self-dual spacetime in loop
  quantum gravity},'' \href{http://dx.doi.org/10.1103/PhysRevD.102.124042}{{\em
  Phys. Rev. D} {\bfseries 102} (2020) 124042},
  \href{http://arxiv.org/abs/2008.08704}{{\ttfamily arXiv:2008.08704 [gr-qc]}}.

\bibitem{Liu:2022qiz}
Y.-L. Liu, Z.-Q. Feng, and X.-D. Zhang, ``{Solar system constraints of a
  polymer black hole in loop quantum gravity},''
  \href{http://dx.doi.org/10.1103/PhysRevD.105.084068}{{\em Phys. Rev. D}
  {\bfseries 105} (2022) 084068},
  \href{http://arxiv.org/abs/2201.10202}{{\ttfamily arXiv:2201.10202 [gr-qc]}}.

\bibitem{Chen:2023bao}
R.-T. Chen, S.~Li, L.-G. Zhu, and J.-P. Wu, ``{Constraints from Solar System
  tests on a covariant loop quantum black hole},''
  \href{http://dx.doi.org/10.1103/PhysRevD.109.024010}{{\em Phys. Rev. D}
  {\bfseries 109} no.~2, (2024) 024010},
  \href{http://arxiv.org/abs/2311.12270}{{\ttfamily arXiv:2311.12270 [gr-qc]}}.

\bibitem{Zhu:2024oxz}
T.~Zhu, H.~K. Nguyen, M.~Azreg-A\"\i{}nou, and M.~Jamil, ``{Observational tests
  of asymptotically flat ${{\mathcal {R}}}^{2}$ spacetimes},''
  \href{http://dx.doi.org/10.1140/epjc/s10052-024-12610-2}{{\em Eur. Phys. J.
  C} {\bfseries 84} no.~3, (2024) 330},
  \href{http://arxiv.org/abs/2402.16922}{{\ttfamily arXiv:2402.16922 [gr-qc]}}.

\bibitem{Zhang:2024khj}
C.~Zhang, J.~Lewandowski, Y.~Ma, and J.~Yang, ``{Black Holes and Covariance in
  Effective Quantum Gravity},''
  \href{http://arxiv.org/abs/2407.10168}{{\ttfamily arXiv:2407.10168 [gr-qc]}}.

\bibitem{Meissner:2004ju}
K.~A. Meissner, ``{Black hole entropy in loop quantum gravity},''
  \href{http://dx.doi.org/10.1088/0264-9381/21/22/015}{{\em Class. Quant.
  Grav.} {\bfseries 21} (2004) 5245--5252},
  \href{http://arxiv.org/abs/gr-qc/0407052}{{\ttfamily arXiv:gr-qc/0407052}}.

\bibitem{Domagala:2004jt}
M.~Domagala and J.~Lewandowski, ``{Black hole entropy from quantum geometry},''
  \href{http://dx.doi.org/10.1088/0264-9381/21/22/014}{{\em Class. Quant.
  Grav.} {\bfseries 21} (2004) 5233--5244},
  \href{http://arxiv.org/abs/gr-qc/0407051}{{\ttfamily arXiv:gr-qc/0407051}}.

\bibitem{Munch:2021oqn}
J.~M\"unch, ``{Causal structure of a recent loop quantum gravity black hole
  collapse model},'' \href{http://dx.doi.org/10.1103/PhysRevD.104.046019}{{\em
  Phys. Rev. D} {\bfseries 104} no.~4, (2021) 046019},
  \href{http://arxiv.org/abs/2103.17112}{{\ttfamily arXiv:2103.17112 [gr-qc]}}.

\bibitem{Lewandowski:2022zce}
J.~Lewandowski, Y.~Ma, J.~Yang, and C.~Zhang, ``{Quantum Oppenheimer-Snyder and
  Swiss Cheese Models},''
  \href{http://dx.doi.org/10.1103/PhysRevLett.130.101501}{{\em Phys. Rev.
  Lett.} {\bfseries 130} no.~10, (2023) 101501},
  \href{http://arxiv.org/abs/2210.02253}{{\ttfamily arXiv:2210.02253 [gr-qc]}}.

\bibitem{lambert2011improved}
S.~Lambert and C.~Le~Poncin-Lafitte, ``Improved determination of $\gamma$ by
  vlbi,'' {\em Astronomy and Astrophysics} {\bfseries 529} (2011) A70.

\bibitem{will2018theory}
C.~M. Will, {\em Theory and experiment in gravitational physics}.
\newblock Cambridge university press, 2018.

\bibitem{Weinberg:1972kfs}
S.~Weinberg, {\em {Gravitation and Cosmology}: {Principles and Applications of
  the General Theory of Relativity}}.
\newblock John Wiley and Sons, New York, 1972.

\bibitem{1992Relativistic}
B.~Bertotti and G.~Giampieri, ``Relativistic effects for doppler measurements
  near solar conjunction,'' {\em Classical and Quantum Gravity} {\bfseries 9}
  no.~3, (1992) 777.

\bibitem{1993Doppler}
B.~Bertotti, G.~Comoretto, and L.~Iess, ``Doppler tracking of spacecraft with
  multi-frequency links,'' {\em Astronomy and Astrophysics} {\bfseries 269}
  no.~2, (1993) 608--616.

\bibitem{Deng:2015sua}
X.-M. Deng and Y.~Xie, ``{Improved upper bounds on Kaluza\textendash{}Klein
  gravity with current Solar System experiments and observations},''
  \href{http://dx.doi.org/10.1140/epjc/s10052-015-3771-4}{{\em Eur. Phys. J. C}
  {\bfseries 75} no.~11, (2015) 539},
  \href{http://arxiv.org/abs/1510.02946}{{\ttfamily arXiv:1510.02946 [gr-qc]}}.

\bibitem{2000Doppler}
L.~Iess, G.~Giampieri, J.~D. Anderson, and B.~Bertotti, ``Doppler measurement
  of the solar gravitational deflection,'' {\em Classical and Quantum Gravity}
  {\bfseries 16} no.~5, (2000) .

\bibitem{Lucchesi:2010zzb}
D.~M. Lucchesi and R.~Peron, ``{Accurate Measurement in the Field of the Earth
  of the General-Relativistic Precession of the LAGEOS II Pericenter and New
  Constraints on Non-Newtonian Gravity},''
  \href{http://dx.doi.org/10.1103/PhysRevLett.105.231103}{{\em Phys. Rev.
  Lett.} {\bfseries 105} (2010) 231103},
  \href{http://arxiv.org/abs/1106.2905}{{\ttfamily arXiv:1106.2905 [gr-qc]}}.

\bibitem{Do:2019txf}
T.~Do {\em et~al.}, ``{Relativistic redshift of the star S0-2 orbiting the
  Galactic center supermassive black hole},''
  \href{http://dx.doi.org/10.1126/science.aav8137}{{\em Science} {\bfseries
  365} no.~6454, (2019) 664--668},
  \href{http://arxiv.org/abs/1907.10731}{{\ttfamily arXiv:1907.10731
  [astro-ph.GA]}}.

\bibitem{GRAVITY:2018ofz}
{\bfseries GRAVITY} Collaboration, R.~Abuter {\em et~al.}, ``{Detection of the
  gravitational redshift in the orbit of the star S2 near the Galactic centre
  massive black hole},''
  \href{http://dx.doi.org/10.1051/0004-6361/201833718}{{\em Astron. Astrophys.}
  {\bfseries 615} (2018) L15},
  \href{http://arxiv.org/abs/1807.09409}{{\ttfamily arXiv:1807.09409
  [astro-ph.GA]}}.

\bibitem{iorio2024general}
L.~Iorio, {\em General Post-Newtonian Orbital Effects From Earth's Satellites
  to the Galactic Center}.
\newblock 2024.

\bibitem{Benczik:2002tt}
S.~Benczik, L.~N. Chang, D.~Minic, N.~Okamura, S.~Rayyan, and T.~Takeuchi,
  ``{Short distance versus long distance physics: The Classical limit of the
  minimal length uncertainty relation},''
  \href{http://dx.doi.org/10.1103/PhysRevD.66.026003}{{\em Phys. Rev. D}
  {\bfseries 66} (2002) 026003},
  \href{http://arxiv.org/abs/hep-th/0204049}{{\ttfamily arXiv:hep-th/0204049}}.

\bibitem{Heidari:2024bkm}
N.~Heidari, A.~A. Ara\'ujo~Filho, R.~C. Pantig, and A.~\"Ovg\"un,
  ``{Absorption, scattering, geodesics, shadows and lensing phenomena of black
  holes in effective quantum gravity},''
  \href{http://dx.doi.org/10.1016/j.dark.2025.101815}{{\em Phys. Dark Univ.}
  {\bfseries 47} (2025) 101815},
  \href{http://arxiv.org/abs/2410.08246}{{\ttfamily arXiv:2410.08246 [gr-qc]}}.

\bibitem{Wang:2024iwt}
Y.~Wang, A.~Vachher, Q.~Wu, T.~Zhu, and S.~G. Ghosh, ``{Strong gravitational
  lensing by static black holes in effective quantum gravity},''
  \href{http://dx.doi.org/10.1140/epjc/s10052-025-13970-z}{{\em Eur. Phys. J.
  C} {\bfseries 85} no.~3, (2025) 302},
  \href{http://arxiv.org/abs/2410.12382}{{\ttfamily arXiv:2410.12382
  [astro-ph.CO]}}.

\bibitem{Gaia:2016zol}
{\bfseries Gaia} Collaboration, T.~Prusti {\em et~al.}, ``{The Gaia Mission},''
  \href{http://dx.doi.org/10.1051/0004-6361/201629272}{{\em Astron. Astrophys.}
  {\bfseries 595} no.~Gaia Data Release 1, (2016) A1},
  \href{http://arxiv.org/abs/1609.04153}{{\ttfamily arXiv:1609.04153
  [astro-ph.IM]}}.

\bibitem{Will:2018mcj}
C.~M. Will, ``{New General Relativistic Contribution to
  Mercury\textquoteright{}s Perihelion Advance},''
  \href{http://dx.doi.org/10.1103/PhysRevLett.120.191101}{{\em Phys. Rev.
  Lett.} {\bfseries 120} no.~19, (2018) 191101},
  \href{http://arxiv.org/abs/1802.05304}{{\ttfamily arXiv:1802.05304 [gr-qc]}}.

\bibitem{Genova:2021rlo}
A.~Genova {\em et~al.}, ``{Geodesy, Geophysics and Fundamental Physics
  Investigations of the BepiColombo Mission},''
  \href{http://dx.doi.org/10.1007/s11214-021-00808-9}{{\em Space Sci. Rev.}
  {\bfseries 217} no.~2, (2021) 31}.

\bibitem{DeMarchi:2019lei}
F.~De~Marchi and G.~Cascioli, ``{Testing General Relativity in the Solar
  System: present and future perspectives},''
  \href{http://dx.doi.org/10.1088/1361-6382/ab6ae0}{{\em Class. Quant. Grav.}
  {\bfseries 37} no.~9, (2020) 095007},
  \href{http://arxiv.org/abs/1911.05561}{{\ttfamily arXiv:1911.05561 [gr-qc]}}.

\bibitem{LISA:2022yao}
{\bfseries LISA} Collaboration, P.~A. Seoane {\em et~al.}, ``{Astrophysics with
  the Laser Interferometer Space Antenna},''
  \href{http://dx.doi.org/10.1007/s41114-022-00041-y}{{\em Living Rev. Rel.}
  {\bfseries 26} no.~1, (2023) 2},
  \href{http://arxiv.org/abs/2203.06016}{{\ttfamily arXiv:2203.06016 [gr-qc]}}.

\bibitem{ET:2019dnz}
{\bfseries ET} Collaboration, M.~Maggiore {\em et~al.}, ``{Science Case for the
  Einstein Telescope},''
  \href{http://dx.doi.org/10.1088/1475-7516/2020/03/050}{{\em JCAP} {\bfseries
  03} (2020) 050}, \href{http://arxiv.org/abs/1912.02622}{{\ttfamily
  arXiv:1912.02622 [astro-ph.CO]}}.

\bibitem{Battista:2017xlm}
E.~Battista, A.~Tartaglia, G.~Esposito, D.~Lucchesi, M.~L. Ruggiero, P.~Valko,
  S.~Dell'Agnello, L.~Di~Fiore, J.~Simo, and A.~Grado, ``{Quantum time delay in
  the gravitational field of a rotating mass},''
  \href{http://dx.doi.org/10.1088/1361-6382/aa7f11}{{\em Class. Quant. Grav.}
  {\bfseries 34} no.~16, (2017) 165008},
  \href{http://arxiv.org/abs/1703.08095}{{\ttfamily arXiv:1703.08095 [gr-qc]}}.

\bibitem{Battista:2014oba}
E.~Battista and G.~Esposito, ``{Restricted three-body problem in
  effective-field-theory models of gravity},''
  \href{http://dx.doi.org/10.1103/PhysRevD.89.084030}{{\em Phys. Rev. D}
  {\bfseries 89} no.~8, (2014) 084030},
  \href{http://arxiv.org/abs/1402.2931}{{\ttfamily arXiv:1402.2931 [gr-qc]}}.

\bibitem{Battista:2014ija}
E.~Battista and G.~Esposito, ``{Full three-body problem in
  effective-field-theory models of gravity},''
  \href{http://dx.doi.org/10.1103/PhysRevD.90.084010}{{\em Phys. Rev. D}
  {\bfseries 90} no.~8, (2014) 084010},
  \href{http://arxiv.org/abs/1407.3545}{{\ttfamily arXiv:1407.3545 [gr-qc]}}.
  [Erratum: Phys.Rev.D 93, 049901 (2016)].

\bibitem{Battista:2015zta}
E.~Battista, S.~Dell'Agnello, G.~Esposito, and J.~Simo, ``{Quantum effects on
  Lagrangian points and displaced periodic orbits in the Earth-Moon system},''
  \href{http://dx.doi.org/10.1103/PhysRevD.91.084041}{{\em Phys. Rev. D}
  {\bfseries 91} (2015) 084041},
  \href{http://arxiv.org/abs/1501.02723}{{\ttfamily arXiv:1501.02723 [gr-qc]}}.
  [Erratum: Phys.Rev.D 93, 049902 (2016)].

\bibitem{Battista:2015wwa}
E.~Battista, S.~Dell'Agnello, G.~Esposito, L.~Di~Fiore, J.~Simo, and A.~Grado,
  ``{Earth-moon Lagrangian points as a test bed for general relativity and
  effective field theories of gravity},''
  \href{http://dx.doi.org/10.1103/PhysRevD.92.064045}{{\em Phys. Rev. D}
  {\bfseries 92} (2015) 064045},
  \href{http://arxiv.org/abs/1507.02902}{{\ttfamily arXiv:1507.02902 [gr-qc]}}.
  [Erratum: Phys.Rev.D 93, 109904 (2016)].

\bibitem{Tartaglia:2018bjc}
A.~Tartaglia, G.~Esposito, E.~Battista, S.~Dell'Agnello, and B.~Wang,
  ``{Looking for a new test of general relativity in the solar system},''
  \href{http://dx.doi.org/10.1142/S0217732318501365}{{\em Mod. Phys. Lett. A}
  {\bfseries 33} no.~24, (2018) 1850136},
  \href{http://arxiv.org/abs/1801.07236}{{\ttfamily arXiv:1801.07236 [gr-qc]}}.

\bibitem{Battista:2020qqp}
E.~Battista, G.~Esposito, and A.~Tartaglia, ``{An effective-gravity perspective
  on the Sun\textendash{}Jupiter\textendash{}comet three-body system},''
  \href{http://dx.doi.org/10.1142/S0219887820501686}{{\em Int. J. Geom. Meth.
  Mod. Phys.} {\bfseries 17} no.~11, (2020) 2050168},
  \href{http://arxiv.org/abs/1910.14551}{{\ttfamily arXiv:1910.14551 [gr-qc]}}.

\end{thebibliography}\endgroup
\end{document}